\documentstyle[pre,aps,psfig,twocolumn]{revtex}

\begin{document}
\twocolumn[\hsize\textwidth\columnwidth\hsize\csname
@twocolumnfalse\endcsname

\title{Cellular Models for River Networks}
\author{Guido Caldarelli}
\address{INFM, Sezione di Roma1, Dip. Fisica, Universit\`a di Roma 
``La Sapienza", P.le A. Moro 2, 00185 Roma Italy} 
\maketitle
%%%%%%%%%%%%%%%%%%%%% Begin Abstract %%%%%%%%%%%%%%%%%%%%%%%%
\begin{abstract}
A cellular model introduced for the evolution of 
the fluvial landscape 
is revisited using extensive numerical and scaling analyses. 
The basic network shapes and their recurrence  
especially in the aggregation structure are then addressed. 
The roles of boundary and initial conditions are carefully analyzed  
as well as the key effect of quenched disorder embedded in 
random pinning of the landscape surface. 
It is found that the above features strongly affect the scaling 
behavior of key morphological quantities.
In particular, we 
conclude that randomly pinned regions (whose structural disorder 
bears much physical meaning mimicking uneven landscape-forming rainfall 
events, geological diversity or heterogeneity in surficial properties like   
vegetation, soil cover or type) play a key role for the robust emergence 
of aggregation patterns bearing much resemblance to real river networks. 
\end{abstract}
%%%%%%%%%%%%%%%%%%%%%% End Abstract %%%%%%%%%%%%%%%%%%%%%%%%%%%%%%
]
\narrowtext

\section{Introduction}
Through experimental studies, it has become evident in the past few years 
that the geometrical and topological structures of  
river basins are characterized by the absence of a single well-defined
length scale.  
This is reflected in the appearance of
power laws in the distribution of several quantities,  
chiefly total contributing area at a point 
\cite{RI92a} and stream lengths 
\cite{T88,LB89,M94,AR96},  and  
by the clear experimental assessment of scaling properties (yielding 
either self-similarity or self-affinity \cite{F88}) 
for many geometrical descriptors of the river basin 
\cite{BM83,AM96,RI96}. 
The discovery of the general underlying mechanisms 
yielding scale-free features 
is the present theoretical challenge.

The network associated to a given natural terrain pertaining 
to a river basin can be 
experimentally analyzed
by using the so-called Digital Elevation Map (DEM) technique 
\cite{T88,RI96,B86,D92,M88,M92} 
which allows to determine the average height of areas (pixels) of the 
order $10^{-2} \mbox{Km}^2 $. Thus a fluvial basin is represented in a 
objective manner often over 4 log scales of linear size.   
Lower bounds are imposed by channel initiation processes at O($10-100$) m.  
Crossovers of geological nature provide altered 
aggregation processes and thus an upper cutoff, usually   
beyond scales of O($10^5$ - $10^6$) m. Thus the observational evidence   
yields a much reliable framework over many scales for comparison 
with dynamical models aimed at the origin of scale-free features. 

Much interest has been recently attracted by landscape evolution models. 
Chief among those are the detailed deterministic models which 
address the description of the detailed dynamics acting on the landscapes 
\cite{H94}. The reductionist approach, where  
a precise description of the details of the dynamics is sought, 
is successful, and much interesting, in the pursuit 
of the description   
and the classification of landforms. 
Nevertheless, as standard in critical phenomena, the 
mechanism producing scale-free structures is expected to depend 
only on a few key features common to all the 
networks rather than on the details of the particular system under 
study. Hence in this work, centered on the dynamic origin of fractal 
river networks, we follow 
a nonreductionist approach based on the simplest possible,  
parameter-free models   
capable of allowing the emergence of complexity. 

A flowrate unit is associated with each pixel
and the flow contributing to any pixel 
follows the steepest descent path through drainage directions 
whose collection defines the planar structure under consideration. 
The resulting network
is therefore the two-dimensional projection of the three-dimensional
tree-like structure of the steepest descent paths draining a given basin.
The planar patterns of network aggregation are obtained by employing 
the cellular model for the 
evolution of a fluvial landscape 
originally introduced by \cite{AR93} and further studied in \cite{C97}
It is aimed at describing in a crucially simple manner the sole fluvial 
component of landscape evolution.  Although such 
component must be coupled to other - chiefly hillslope - transport 
processes to yield a 
comprehensive dynamical description \cite{D92,H94,W91,Hetal94,R94}, 
it rules the planar 
imprinting of the network. 
Hence the detailed study of the model is 
deemed significant. 

Starting from a three dimensional landscape, evolution occurs according
to a threshold dynamics similar to the one proposed in self organized 
critical (SOC) models \cite{B87}. 
The main idea of the erosion dynamics is that whenever the local
shear stress exceeds a given threshold, erosion starts an 
`avalanche' and a related 
rearrangement of the network patterns takes place.  
The model of self-organizing fluvial structures 
may be seen as a modification of the sandpile model 
developed as a paradigm of the dynamics of open, 
dissipative systems with many degrees of freedom.   
It may be thought of as belonging to the set of models 
in which the threshold 
for activity, say $\tau_c$,   
rather than being a constant value,   
depends on non-local properties of the self-organizing structure. 
In the fluvial case, the non-local character of the threshold value 
follows from the fact that the threshold at the arbitrary 
site equals a shear stress, i.e. 
$\tau \propto \nabla h \sqrt{a}$, where $h$ is the local landscape elevation 
and $a$ is total contributing area surrogating total flow collected 
from a distributed rainfall event. As such, 
the exceedence of $\tau_c$ depends not only on 
local conditions (i.e., a 
critical value of $\nabla h$), but also on non-local conditions 
defined by 
the contributing area $a$ computed through drainage directions, 
i.e. it depends on the entire state of the system which is 
self-organizing. 
Notice that the physical rationale for the nonlocal dependence 
lies in the fact 
that the system is open, i.e. injected from outside, allowing 
flow rates to be proportional to total contributing drainage area.  
The long-range nature of the threshold dynamics 
tends to hide from the observer the temporal fluctuations which take 
place in the evolutionary time scale. 
In this sense  
the above model was classified \cite{AR93} as one of spatial 
self-organized criticality.  

Whether or not river self-organization qualifies as a more 
general framework of self-organized criticality remains to be seen. 
If SOC must necessarily refer to 
the occurrence of a critical state in the sense of critical phenomena,  
where a small local perturbation can cause a significant 
change in the configuration of the whole system and thus the 
system  
shows both spatial and temporal scaling, then  
the time dynamics should be specifically considered. One way to 
do this is through  
the oscillation of the threshold in time, i.e. 
$\tau_c(t)$, simulating climatic fluctuations  
(see \cite{AR95}), 
through which indeed temporal evolution appear, 
or through perturbations of random location and strength in the 
evolution of the landscape. This is also true, as we will 
discuss later, if the landscape-forming rainfall events are described as 
nonuniform in space, leading to patches of activity randomly scattered 
spatially (in such a case the outflow response of the system  
becomes a $1/f$ signal). 
However, regardless of any additional features, 
we believe that the central 
scope of SOC  
is the dynamic explanation of 
the growth of fractal structures of the type appearing in nature, i.e. 
the physics of fractals. As such we feel that our classification 
of the model as a particular case of SOC is a suitable one regardless 
of the description of the embedded temporal activity because the 
system always reaches a fractal state. 
Moreover, questioning on this basis \cite{S96} 
the self-organized critical nature of the  
model by Ref. \cite{AR93} is irrelevant  
because it has been shown on thermodynamics grounds that scaling properties 
of energy and entropy yield limit states 
which, depending on the constraints, are temporally frozen or active 
\cite{AR96}. 
Furthermore, ref. \cite{RI96} shows that 
optimal states like the ones dynamically accessed by the above model 
may exist in temporally active states precisely at the edge of a chaotic 
behaviour. 

A question, indeed more interesting than the semantics of 
SOC, is whether 
the constraints in the model may be relaxed to produce 
a 'hot' fluvial landscape more closely resembling an ordinary sandpile.   
This question is addressed in \cite{AR96} and in more detatil in \cite{RI96}. 

Our main goal is twofold. On one hand we will extend previous investigations
both in accuracy and in statistics by performing simulations at much
larger scales. On the other hand, we will consider important issues such
as the effect of the boundaries and of the initial 
conditions bearing much significance on geologic influences.
In particular we will show
that both the aforementioned effects play an important role in 
the results previously obtained. We also study the effects of 
disorder, say through the presence of small, uncorrelated 
inhomogeneities in the initial conditions,  
in particular with regard to the robustness to single/multiple outlet 
arrangements. 

The paper is organized as follows. In the first 
section 1 the
model is recalled. Section 2 presents the results with emphasis 
on scaling analyses, while the following 
section focuses on the important 
effects of quenched pinning on the system.
A set of conclusions closes then the paper.

\section{The model}

We consider a lattice model of a real landscape.
Let $h_x$ be the height of the landscape associated to  every site $x$ 
of a square lattice of size $L \times L$. 
The lattice is tilted at an angle $\theta$ with respect to a given 
axis to mimic the effects of gravity. 
Two possibilities will be analyzed:
\begin{itemize}
\item [1] All the sites on the lowest side (kept at height $h=0$)
are possible outlets (i.e. the multiple outlet arrangement) 
of an ensemble of rivers 
which are competing to drain the whole $L \times L$ basin;
\item [2] Only one site is kept at $h =0$ 
and it is the outlet of a single river in the $L \times L$ 
basin.
\end{itemize}

In addition, for both the above cases, two types of initial 
conditions will be considered: (a) a regular initial landscape, e.g. 
flat, and (b) an irregular surface obtained by superposing to a smooth  
sloping surface a suitable noise. 

Each site collects an unit amount of water from 
a distributed injection (here a constant rainfall rate as in the original 
approach) in addition 
to the flow which drains into it from the upstream sites.
A unit of water mass is assigned to each pixel of drainage area
so that the total area drained into a site is also a measure
of the total water mass collected at that site.
From each site water flows to one 
of the eight sites, four nearest neighbors 
and four next nearest neighbors, having the lowest height 
(i.e. the steepest descent path). 
We shall
indicate all these eight neighbouring sites as nearest-neighbours ($nn$).
This construction allows the assignment of drainage directions to an 
arbitrary landscape. The drained area $a_x$ is associated to each site $x$ 
according to the equation
\begin{eqnarray}
a_x&=&\sum_{y(x)} a_y +1
\label{eq:1}
\end{eqnarray}
where the sum runs over the subset $y(x) \in nn(x)$ of neighbor sites 
whose area is actually drained by $x$.
The second term in eq.(\ref{eq:1}) represents the uniform injection.

Also, the (up)stream-length $l_x$ from site
$x$ to the source is computed according to the following procedure. 
At a given
site $x$ the areas of all $nn(x)$ of that sites are checked. 
Following the ordinary meaning of downstream and upstream sites 
(i.e. downstream is the site one finds following the river to the outlet,
upstream is the site following which one reach the source from which the
largest incoming river enters the site.)
Following Ref.\cite{AM96}, the $nn$ with largest
value leads to the outlet and is defined to be a downstream
site. The $nn$ with the second largest value indicates the longest path
toward the source and is defined to be the upstream site. The sum
of all the upstream sites from site $x$ to the source is $l_x$.
The downstream length could be defined through an analogou procedure. 
Experimental measures are available for both 
$a_x$ and $l_x$ \cite{RI96}.

The time evolution of the model follows the following steps:
\begin{itemize}
\item[1] The shear stress $\tau_x$ acting at every site is 
computed according to \cite{AR93}
\begin{equation}
\tau_x = \Delta h_x \sqrt{a_x}
\end{equation}
where $\Delta h_x$ is the local gradient along the drainage direction.
\item[2] If the shear stress at a site exceeds a 
threshold value, $\tau_c$, 
then the corresponding height $h_x$ is reduced (i.e. by erosion) in order 
to decrease the local gradient. The shear stress 
is set just at the threshold value.
This produces a rearrangement of the network followed by
a reupdating of the whole pattern as in step 1.
\item[3] When all sites have shear stress below threshold the system is
in a dynamically steady state. 
Since this situation is not necessarily the most stable, a perturbation
is applied to the network with the aim of increasing the stability 
of a new steady state.
A site is thus chosen at random and its height is increased in such a way 
that no lakes, i.e. sites whose height is lower than that  
of their eight neighbours, are formed. Steps 1 and 2 then follow as before.
\end{itemize}

After a suitable number of the perturbations (step 3), the system reaches
a steady state which is unsensitive to further perturbations and where 
all statistics of the networks are stable.
This resulting state is scale-free,  i.e. 
it is characterized by  
power-law distributions of the physical quantities of interest. 

\section{Results}
\label{sec:results}
\subsection{Landscape evolutions}
\label{subsec:landscape}
Our numerical calculations were carried out on a bidimensional 
square lattice (where each site has eight nearest-neighbours) for sizes 
up to $L=200$
with reflecting boundary conditions in the direction transversal to
the flow and open boundary condition in the parallel one. 
We considered the following initial conditions:

\begin{description}

\item{{\bf Model A}} A comb-like structure with a single outlet 
This was the situation originally studied in \cite{AR93}  
and our results are in agreement with theirs.

\item{{\bf Model B}} An inclined plane with all sites at the bottom of
the plane allowed to be possible outlets. This choice was selected with the
aim of investigating the differences 
arising when arranging the boundary conditions with multiple outlets versus 
single outlet. The former allows for  
competition  for drainage area among rivers. 

\item{{\bf Model C,D}} The two previously considered situations with the
addition of a random, uncorrelated noise 
(whose strenght - i.e. variance - is less than $10\%$ of the average height). 
That is, on top of the height computed according to the rules of model A and
B respectively (a comb-like lattice and an inclined plane), we added a random
$Dh$ that is extracted in the interval $[-<h>/10,<h>/10]$ where $<h>$ represents
the mean altitude of the landscape.

\end{description} 

An average over a few (up to five) configurations was 
taken. This choice, especially when coupled to  
large sizes of the system, proves sufficient for a 
statistical descriptions sought     
in view of the self-averaging
nature of the random perturbation.

In Figure 1 typical landscapes sculpted by the above dynamical
process and the corresponding networks drawn through the 
steepest descent construction are shown for models 
A,B. The same picture for models C,D is 
shown in Figure 2. 

Two features can be grasped from these pictures. First, in the results
of both model A and B there is a strong memory of the initial configuration
despite the fact that the dynamics of the erosion process was 
somewhat expected to
be sufficiently strong to soon lose the imprinting 
of its initial condition. Secondly, 
the single outlet restriction imposed in model A appears to be a 
severe constraint because it increasingly 
affects the wandering of the main river
towards the lowest part
of the basin. Our results suggest that this is indeed the case
for flat initial conditions (A and B) while
for noisy initial conditions boundary effects are of lesser importance. 

\subsection{Area and length exponents}

Let us define $P(a,L)$ and $\Pi(l,L)$ as the exceedence 
(cumulative) probability 
distributions of the drainage area $a$ and stream length
$l$ respectively arising in a domain of linear size $L$.  
The following scaling forms are 
expected to hold
\cite{AM96}: 
\begin{eqnarray}
\label{areas}
P(a,L) &=& a^{1-\tau} F(\frac{a}{L^{1+H}})
\end{eqnarray}
\begin{eqnarray}
\label{lengths}
\Pi(l,L) &=& l^{1-\gamma} G(\frac{l}{L^{d_l}})
\end{eqnarray}
Here $H$ is the Hurst exponent \cite{F88}
and $d_l$ is the stream-length (or chemical distance) fractal exponent. 

As it was already noted \cite{AM96}, for self-affine 
river networks ($H<1$, $d_l=1$) the scaling relations 
relate all exponents in terms of $H$. For self-similar river 
networks ($H=1$, $d_l>1$), the same happens in terms of $d_l$. 

Another important indicator of basin morphology is
the relation between the mean total contributing area 
$a$ and the length of the main stream 
$l_{\mbox{max}} \propto L^{d_l}$ \cite{AM96,R96}, 
which is commonly known as Hack's law \cite{H57}:
\begin{eqnarray} 
\label{hack}
a &\propto & l_{max}^{1/h} \ \  \propto \ \ L^{d_l/h}
\end{eqnarray}
The related exponent has been studied in all simulations. 
A summary of the scaling relations between the various exponents 
involved is reported in Table \ref{table1}. 

Experimental values of $\tau$ and $\gamma$ are available 
from earlier analyses of DTMs from basins of different size, 
geology, exposed lithology, 
climate and vegetation \cite{RI92a,R96}. 
It was observed that, while  
a majority of basins tend to seemingly universal values 
$\tau = 1.43 \pm 0.02$ and $\gamma = 1.8 \pm 0.1$, 
exceptions are observed where altered values are observed 
although always in a concerted  
manner. Since it was suggested \cite{AM96} that scaling laws  
for river networks are related, e.g. $\gamma = 1 + (\tau-1)/h$, it 
was concluded there that no universal exponents are 
expected in nature. Rather, the roles of geology and tectonics 
concert a coordinated 
scaling structure which strives for fractality yet adapted to its 
geological environment. The results of the model described here, 
revisited in the above light, conform to this view. 

The results for the four models A,B,C and D 
for the area distributions are shown in
Figure 3. It is apparent that, due to the pathological 
initial conditions, the scaling behaviour for models A and B is somewhat
more noisy than for models C and D. Figure 4 contains
the collapse plot for all the cases. 
Figure 5 shows the stream-length
distribution for the four models. For this picture the same remarks of 
Figure 3 apply.
In Figure 6 we show the collapse plot
corresponding for the stream-length distributions. 

A summary of the scaling exponents obtained 
is included in Table \ref{table2} where 
we observe a consistent picture of related scaling exponents as 
theoretically expected - see Table \ref{table1}.  

\subsection{Energy Dissipation and Optimal Channel Networks}

During the evolution of the landscape 
we also monitored the change in 
total energy dissipation of the system, 
defined as $E = \sum_x a_x^{0.5}$ (where $x$ spans all sites of the lattice) 
\cite{RI92b,RI92c,AR92}. 
The reason of this name come from the computation at any site of the
gravitational energy lost by the falling of the water.
In any point $x$ one can expect a gravitational energy loss of the
order of $a_x \Delta h_x$ where $\Delta h_x$ represents the local
gradient along the drainage direction.
By using the observed scaling 
$\Delta h_x \propto {a_x}^{0.5}$ one obtains the above formula.
The interest in this quantity comes from the fact that an extensively
class of models known as Optimal Channel Network (OCN) models, assume
this quantity is minimized by natural landscape evolution.
By using this principle OCN describe evolution from random spanning 
graphs  to network more similar to the real ones. 
It is interesting to note that in this model where no 
hypothesis is made on $E$, we still observe an almost monotonical 
decrease of $E$ associated with the  
dynamical evolutions, and a stabilization on 
different plateaus of values of $E$.  
The actual figures for a sample $30 \time 30$ are as follows:
$E$ starts from
an initial value of $7600$ and decreases towards a plateau of 
$6800-6700$ where this monotonic decrease becomes slower ($1\%$ decrease 
in 50000 steps). 
This behavior, also observed in other models \cite{SB96}, 
bears important consequences in the light of the suggested 
connection of fractality and optimality 
\cite{AR96,RI92b,RI92c,AR92}

\section{Geological constraints and Quenched Random Pinning}

This section presents a detailed study on 
the effects on landcape evolution of quenched randomness, simulated 
by a random choice of sites unable to evolve regardless of the threshold 
value developed therein. 
It is found that this form of disorder 
tends to favour aggregation patterns characterized by values of 
$\tau = 1.43 \pm 0.02$ for both models, say, A and B (i.e. with single 
outlets or open boundary conditions). This results suggests that the 
origin of the recurrent values observed in nature could be related 
to the ubiquity of heterogeneity in surface properties characterizing locally  
the critical shear threshold. 

Within the river basin morphological and geological 
constraints play a definite 
role in the dynamical evolution of landforms. 
The effects of quenched constraints, simulating any heterogeneity 
is the distribution of surficial properties affecting erosion 
properties, is to favour some sites 
for the  
flowpaths, thus excluding other sites 
from the capture of the developing network which ultimately shapes 
the evolution process. 
In order to mimic such effects we analyzed the effect
of a random pinning of a small region of the total surface 
(typically $5-9\%$)
where the evolution is frozen, that is, the height is pinned to
its initial value. We find that this constraint tends to favour
aggregation even in the presence of random initial conditions.  

As regards the effect of the pinning, figure 7 (on the left) shows  
a sample whose dimension is $100 \time 100$ with multiple
outlet and random initial noise (model D). Figure 7 (on the right) 
shows the same configuration (evolved from the same initial
conditions) but with a $5\%$ dilution pinning. 
It is evident that some of the smaller streams on the left have 
increased their size thus leading to a bigger aggregation. 

In this case we found $\tau=1.43 \pm 0.02$. Moreover, all the other 
exponents verify the correct scalings predicted in Table I.
Purely for comparison purposes with the prevous results, we also report
the plot of the area distribution in Figure 8 on the left. 
We also found that model D reproduces the same results, 
i.e. $\tau = 1.43 \pm 0.04$ and $\gamma = 1.60 \pm 0.04$.
We are confident that this result at least for model C is quite
robust with respect to changes of the pinning dilution.
In the case of a $9\%$ dilution for a smaller number of simulation
we found quite similar results $\tau=1.44 \pm 0.05$ and a 
cumulative plot of $P(a'>a,L)$ is shown on the right part of 
Fig.8\cite{data}.

This result suggests that the origin of recurrent values observed
in nature could indeed be related to the ubiquity of geological
and morphological constraints in the surface properties
locally characterizing the critical shear stress.

\section{Conclusions}
\label{sec:conclusions}

In this paper we revisited the model originally introduced in 
\cite{AR93} which we extended both in accuracy and goals.
Specifically, we analyzed the stability of the universality
class of the original model with respect to the initial conditions
and to the change from single to multiple outlets.
We found that if one starts with structured initial conditions,
critical exponents are sensible to a change from single to multiple
outlets. On the other hand, upon starting from disordered initial
conditions, we found critical exponents belonging to a new 
class which appears to be robust to the change from single to multiple
outlet. Thus this simple model, under controlled conditions,  
yields somewhat different yet internally 
consistent scale-free fluvial landforms depending  
on the dominant conditions affecting evolution.   

The above results conform to the experimental observation 
\cite{AM96}
suggesting that the relevant scaling exponents for river networks are 
not universal. Rather, the fractal nature of river networks 
adjusts to the constraints imposed by the geological environment in a 
coordinated manner. 
It is interesting to observe that the final state of all 
simulation yields indeed fractal structures, 
as observed in nature, though characterized by 
different aggregation properties. 
The exponents characaterizing these different aggregates, nevertheless
follow in a rather good agreement the scaling relations shown in Table I.

We suggest that the 
lack of robustness in the value of the scaling exponents with respect to 
boundary and  initial conditions is related to the non-local character of the 
shear-based threshold, differently from what is observed in 
classical sandpile models of self-organized criticality 
\cite{B87}.  
The intrinsic interest of the different aggregation properties of 
the steady states of the dynamics is related to their optimality 
with respect of total energy dissipation \cite{AR93}.  
In fact, depending on external conditions, 
the same dynamical process may indeed get trapped in steady state 
configurations yielding 
local minima of the total energy dissipation functional, in what 
we may define as a feasible optimality process \cite{AR96}. 

Finally, we have found that quenched 
disorder, modeled by random pinning, has a profound effect on the robustness  
of the resulting planar patterns by favouring aggregation and by locking  
the planar landforms into modes quite similar to the ones observed in nature. 

The remarkable success obtained by such a simple model in enlightnening
some crucial features of the real basins is promising
for a future success in 
a general characterization of the dynamics of fractal growth.

\acknowledgments Enlightening discussions with J. Banavar, 
F. Colaiori A. Flammini, A. Maritan, and A. Rinaldo are acknowledged. 
Suggestion by A. Giacometti have been particularly helpful.
Author acknowledge support from EU contract FMRXCT980183.

%%%%%%%%%%%%%%%%%%%%%%%%%%%%%%%% Tables %%%%%%%%%%%%%%%%%%%%%%%%%%%%%
\begin{table}
%%%%%%%%%%%%%%%%%%%%%%%%%% TABLE %%%%%%%%%%%%%%%%%%%%%%%%%%%%%%%
% Table 1 %
\begin{tabular}{|l|c|c|}
\multicolumn{1}{l}{Exponent}&
\multicolumn{1}{c}{Self-Similar}&
\multicolumn{1}{c}{Self-Affine} \\
\hline
$\tau$   & $2-d_l$ & $(1+2H)/(1+H)$\\
$\gamma$ & $2/d_l$ & $1+H$\\
$h$      & $d_l/2$ & $1/(1+H)$\\
\hline
\end{tabular}
\caption{ Scaling relations: all the exponents can 
be determined in terms of
$d_l$ in the fractal case and $H$ in the self-affine case.}
\label{table1}
\vspace{1cm}

% Table 2 %
\begin{tabular}{|l|c|c|c|c|}
\multicolumn{1}{l}{}
&\multicolumn{1}{c}{Model A}
&\multicolumn{1}{c}{Model B}
&\multicolumn{1}{c}{Model C} 
&\multicolumn{1}{c}{Model D} \\
\hline
$\tau$ & $1.43 \pm 0.03$ & $1.50 \pm 0.03$&
$1.38 \pm 0.02$ & $1.38 \pm 0.02$\\
$\gamma$& $1.60 \pm 0.05$ &$1.70 \pm 0.05$&$1.60 
\pm 0.02$&$1.60 \pm 0.02$\\
$h_{th}$& $0.72 \pm 0.05$ &$0.71\pm 0.05$&$0.63\pm 0.05$&$0.63\pm 0.05$\\
$h_{ex}$& $0.72\pm 0.05$ &$0.69\pm 0.02$&$0.65\pm 0.02$&$0.65\pm 0.02$\\
\hline
\hline
\end{tabular}
\caption{Data computed for the various computer simulations. Since the 
scaling relationship $h=\frac{\tau-1}{\gamma-1}$ one can also 
compute the theoretical value $h_{ex}$ with the measured one $h_{ex}$. The
agreement of the two is rather good.}
\label{table2}
\end{table}
%%%%%%%%%%%%%%%%%%%%%%%%%%%%%%%%%%%%%%%%%%%%%%%%%%%%%%%%%%%%%%%%%%%%%%%%%%%%%%

%%%%%%%%%%%%%%%%%%%%%%%%%%%%%%%% Figures %%%%%%%%%%%%%%%%%%%%%%%%%%%%%
\begin{figure}
\centerline{\hbox{\psfig{figure=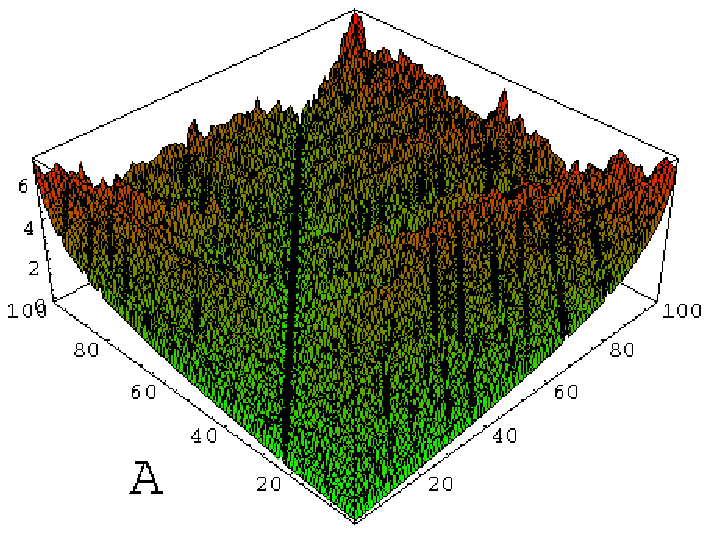,height=3.5cm}
                  \psfig{figure=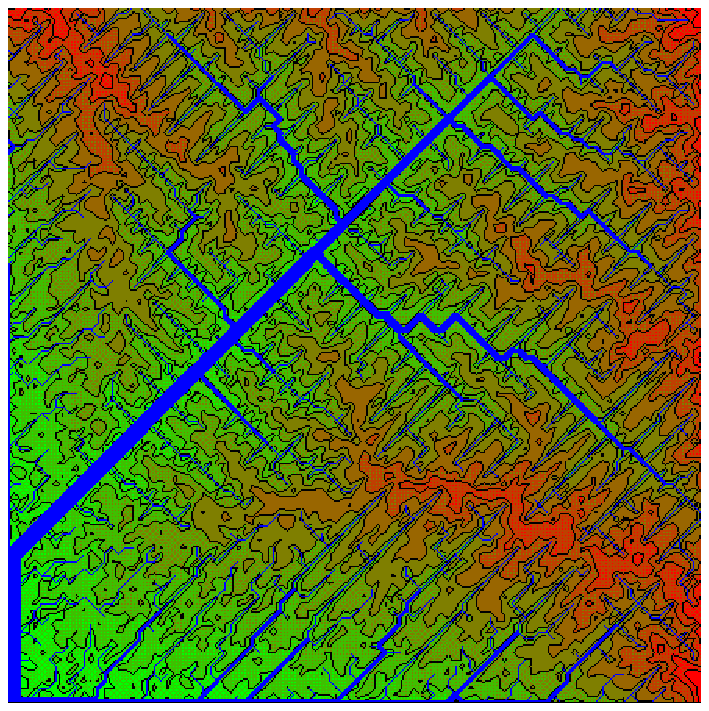,height=3.8cm} }}
\vspace{0.2cm}
\centerline{\hbox{\psfig{figure=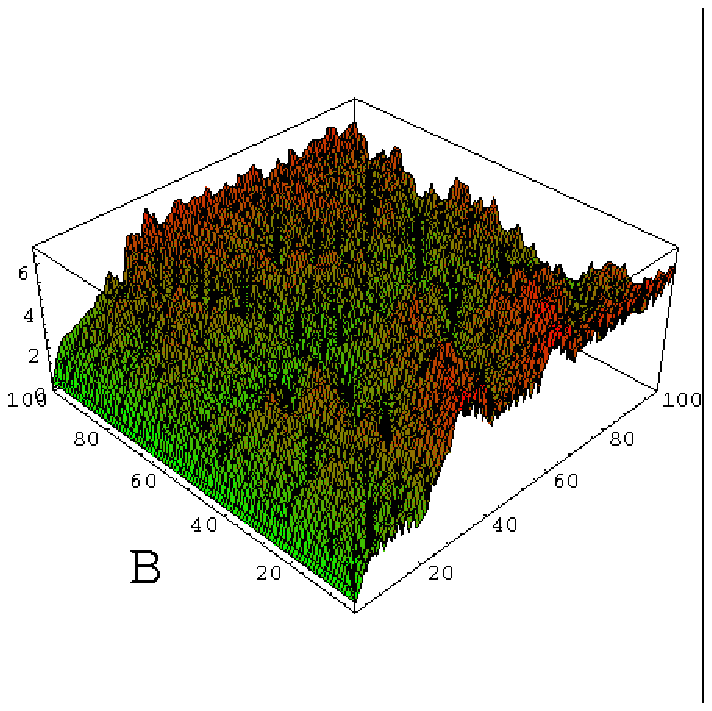,height=3.5cm}
                  \psfig{figure=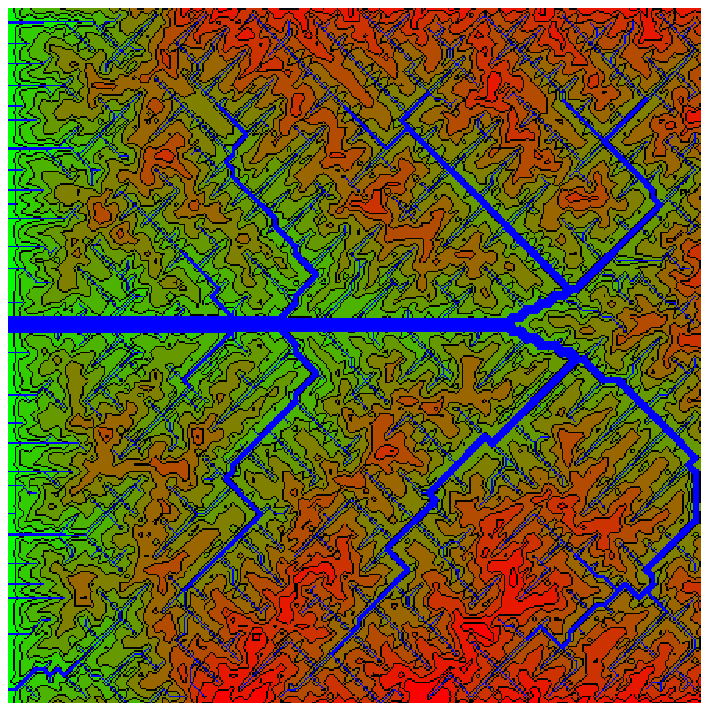,height=3.8cm} }}
\caption{The final landscape (on the left)
and the final network structure (on the right)for Model A and model B.
Models A and B, start from a deterministic initial condition. Model A
has single outlet, model B has multiple outlets.}
\label{fig1}
\end{figure}

%Fig 2%
\begin{figure}
\centerline{\hbox{\psfig{figure=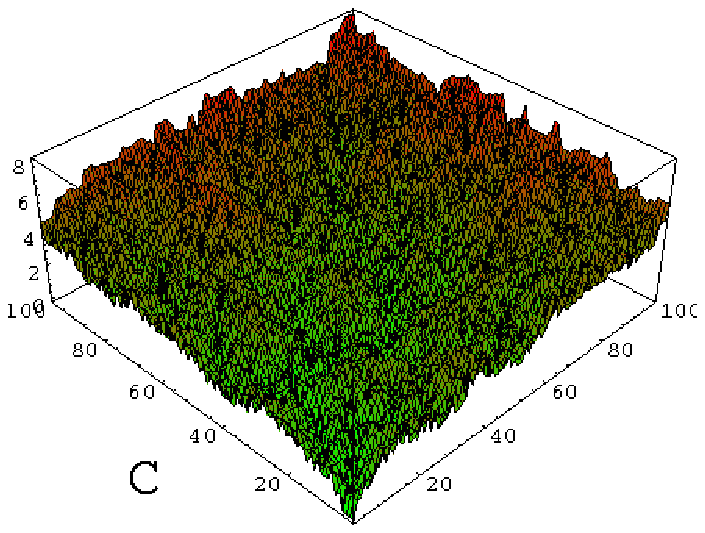,height=3.5cm}
                  \ \ \psfig{figure=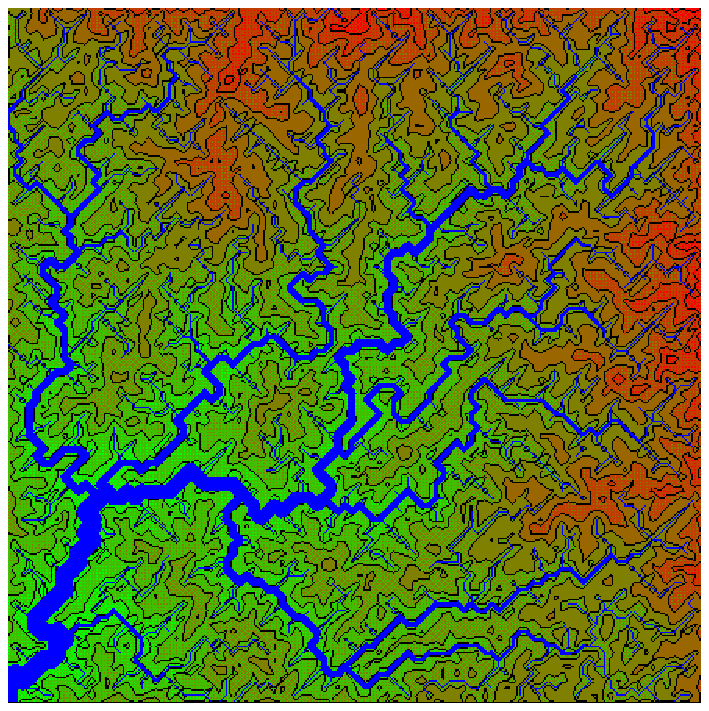,height=3.8cm} }}
\vspace{0.33cm}
\centerline{\hbox{\psfig{figure=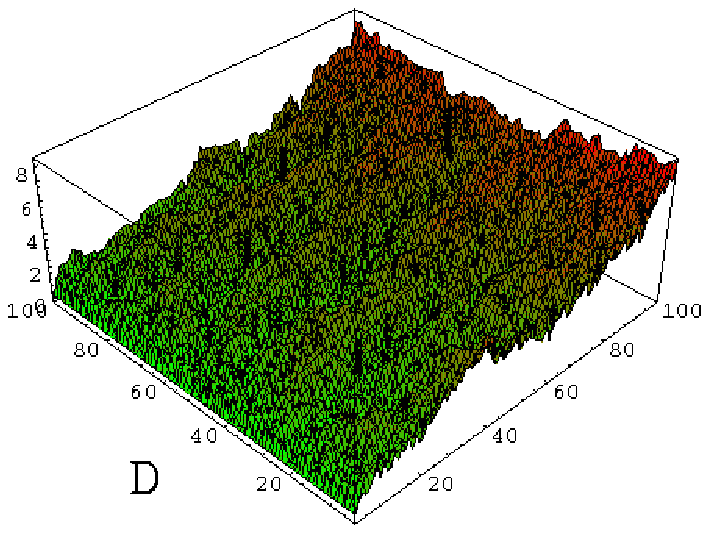,height=3.5cm}
                 \ \ \psfig{figure=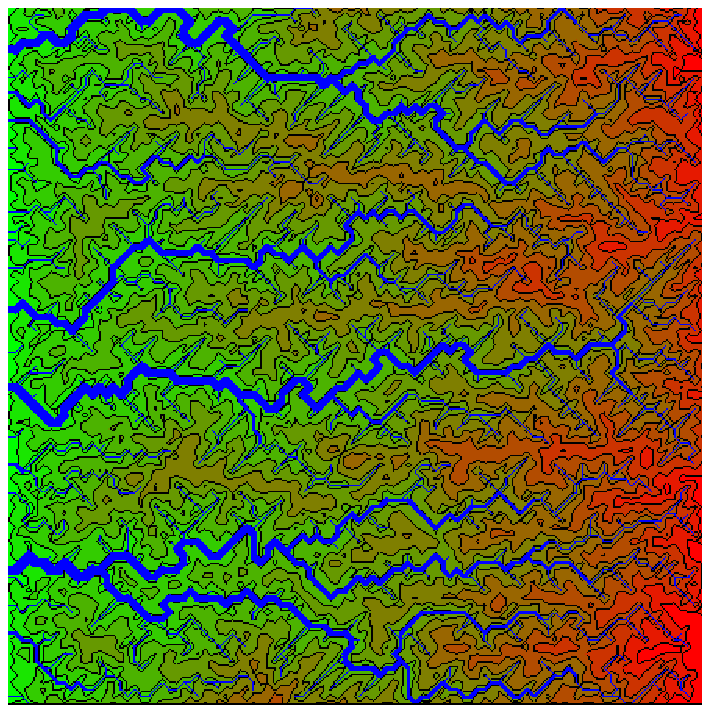,height=3.8cm} }}
\vspace{0.33cm}
\caption{The final landscape (on the left)
and the final network structure (on the right)for Model C and model D.
Models C and D, start from a random initial condition. Model C
has single outlet, model D has multiple outlets.}
\label{fig2}
\end{figure}

%Fig 3%
\begin{figure}
\centerline{\psfig{figure=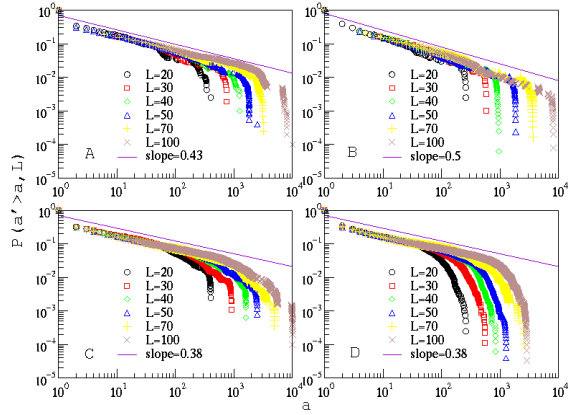,height=5.7cm}}
\caption{Log-Log plot of the area cumulated distribution $P(a,L)$
versus $a$ for model A,B,C and D.
The full line has a slope corresponding
to $\tau=1.43$, $\tau=1.50$, $\tau=1.38$ and $\tau=1.38$ respectively.}
\label{fig3}
\end{figure}

%Fig 4%
\begin{figure}
\centerline{\psfig{figure=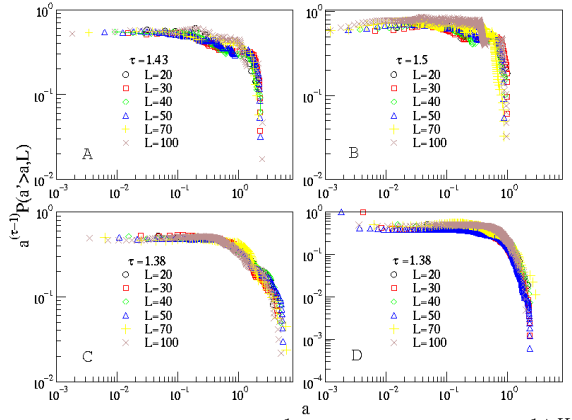,height=5.7cm}}
\caption{Scaling function $a^{1-\tau} P(a,L)$ versus $a/L^{1+H}$
for model C. The used values to obtain the collapse were 
$\tau=1.43,\tau=1.5 \tau=1.38$, and $\tau=1.38$ for Model A,B,C,D 
respectively. $H=0.6$ in all the cases}
\label{fig4}
\end{figure}

%Fig 5%
\begin{figure}
\centerline{\psfig{figure=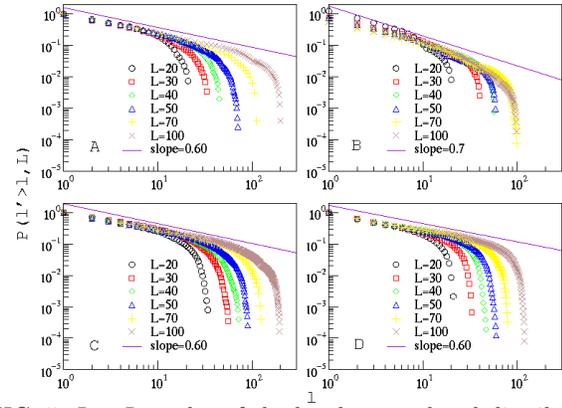,height=5.7cm}}
\caption{Log-Log plot of the lenght cumulated distribution $P(l,L)$
versus $l$ for model A,B,C and D.
The full line has a slope corresponding
to $\gamma=1.6$, $\gamma=1.7$, $\gamma=1.6$ and $\gamma=1.6$ respectively.}
\label{fig5}
\end{figure}

%Fig 6%
\begin{figure}
\centerline{\psfig{figure=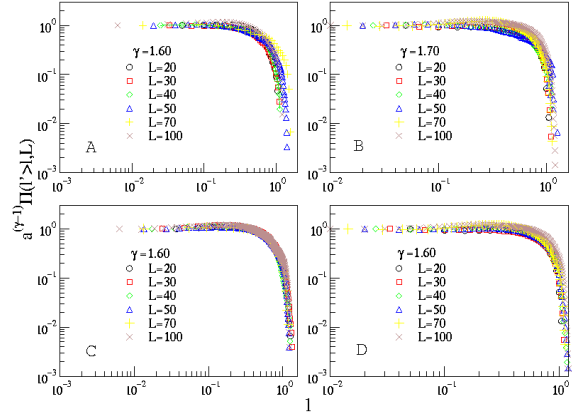,height=5.7cm}}
\caption{Scaling function $l^{1-\gamma} \Pi(l,L)$ versus $l/L^{d_l}$
for models A,B,C,D. The values used to obtain the collapse were 
to $\gamma=1.6$, $\gamma=1.7$, $\gamma=1.6$ and $\gamma=1.6$ respectively.}
\label{fig6}
\end{figure}

%Fig 7%
\begin{figure}
\centerline{\psfig{figure=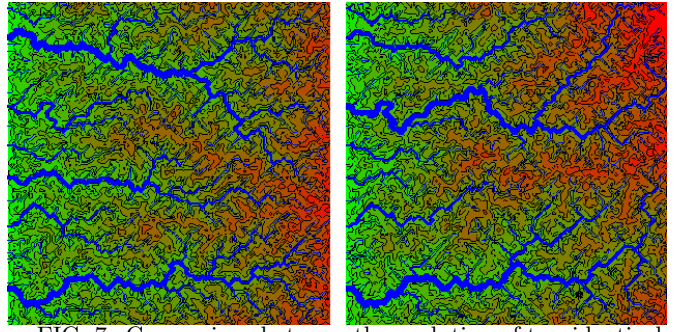,height=4.3cm}}
%\centerline{\hbox{\psfig{figure=fig7a.ps,height=8cm}
%                  \ \ \psfig{figure=fig7b.eps,height=8cm} }}
%\vspace{0.33cm}
\caption{Comparison between the evolution of two identical
initial configurations of model C with size $L=100$ without (left) and with
pinning (right). The pinning diluition was $5\%$.}
\label{fig7}
\end{figure}

%Fig 8%
\begin{figure}
\centerline{\psfig{figure=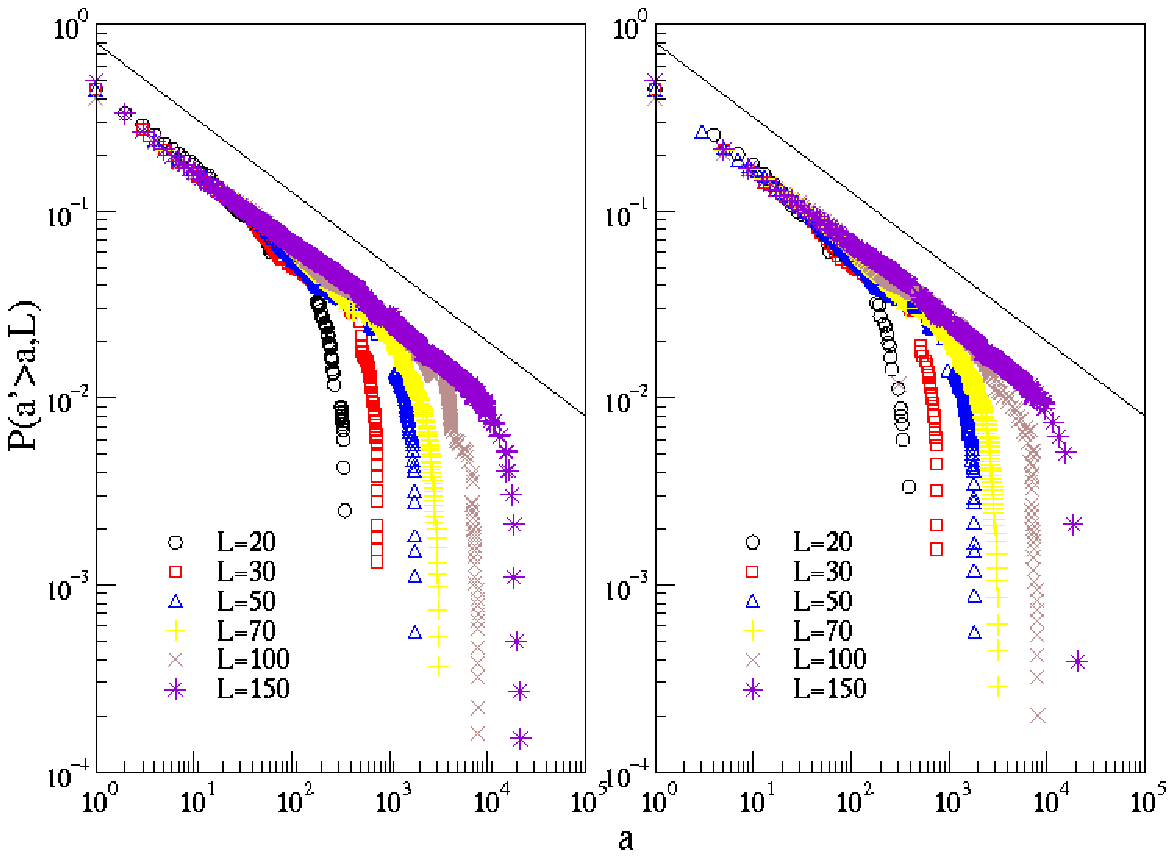,height=5.7cm}}
\caption{Log-Log plot of the area distribution $P(a,L)$ vs. $a$ for model C 
with a $5\%$ dilution(left) and with a $9\%$ dilution (right). The slope of the lines
corresponds respectively to $\tau=1.43$ and $\tau=1.44$.}
\label{fig8}
\end{figure}


\begin{thebibliography}{99}

\bibitem{RI92a}
I. Rodriguez-Iturbe, E. Ijjasz - Vasquez, R.L. Bras and D.G.
Tarboton, 
%Power-law distributions of mass and energy in river basins,
{\it Water Resour. Res.},  28(3), 988-993, (1992).

\bibitem{T88}
D.G. Tarboton, R. L. Bras and I. Rodriguez-Iturbe, 
%The fractal nature of river networks, 
{\it Water Resour. Res.}, 24, 1317-1322, (1988).

\bibitem{LB89}
P. La Barbera, and R. Rosso, 
%On the fractal dimension of stream networks, 
{\it Water Resour. Res.}, 25(4), 735-741, (1989).

\bibitem{M94}
M. Marani, A. Rinaldo, R. Rigon and I. Rodriguez-Iturbe, 
%Geomorphological width functions and the random cascade, 
{\it Geophys. Res. Letters}, 21(19), 2123-2126, (1994).

\bibitem{AR96}
A. Rinaldo,  A. Maritan, A. Flammini, F. Colaiori, R. Rigon, I. 
Rodriguez-Iturbe and J.R. Banavar, 
%Thermodynamics of fractal networks, 
{\it Phys. Rev. Lett.}, 76, 3364-3368, (1996).

\bibitem{F88}
J. Feder, {\it Fractals}, Plenum, New York, (1988).

\bibitem{BM83}
B.B. Mandelbrot, {\it The Fractal Geometry of Nature},
Freeman, New York, (1983).

\bibitem{AM96}
A. Maritan, A. Rinaldo, R. Rigon, A. Giacometti
and I. Rodriguez-Iturbe, 
%Scaling laws for river networks, 
{\it Phys. Rev. E}, {\bf 53}, 1510-1515, (1996).  

\bibitem{RI96}
I. Rodriguez-Iturbe and A. Rinaldo, {\it Fractal River Basins: 
Chance and Self-Organization},
Cambridge University Press, (1996). 

\bibitem{B86}
L. Band, 
%Topographic partition of watersheds with digital elevation models, 
{\it Water Resour. Res.}, 22(1), 15-24, (1986).

\bibitem{D92}
W.E. Dietrich, C. J. Wilson, D.R. Montgomery, J. Mc Kean, 
%Erosion thresholds and land surface morphology,  
{\it J. Geol.}, {\bf 20}, 675-679, (1992).

\bibitem{M88}
D.R. Montgomery, and W.E. Dietrich,  
%Where do channels begin?, 
{\it Nature}, {\bf 336}, 232-234, (1988). 

\bibitem{M92}
D.R. Montgomery, and W.E. Dietrich,  
%Channel initiation and the problem of landscape scale, 
{\it Science}, {\bf 255}, 826-830, (1992).

\bibitem{H94}
A.D. Howard, 
%A detachment-limited model of drainage basin evolution, 
{\it Water Resour. Res.}, 30(7), 2261-2285, (1994).

\bibitem{AR93}
A. Rinaldo, I. Rodriguez-Iturbe, R. Rigon, E.
Ijjasz-Vasquez and R. L. Bras, 
%Self-organized fractal river networks, 
{\it Phys. Rev. Lett.} {\bf 70}, 822-826, (1993).

\bibitem{C97}
G. Caldarelli, A. Giacometti, A. Maritan, 
I. Rodriguez-Iturbe, A. Rinaldo
{\it Phys. Rev. E} {\bf 55}, R4865-R4868 (1997). 

\bibitem{W91}
G.R. Willgoose, R.L. Bras and I. Rodriguez-Iturbe, 
%A coupled channel network growth and hillslope evolution model, 1, Theory, 
{\it Water Resour. Res.}, 27, 1671-1684, (1991).

\bibitem{Hetal94}
A.D. Howard, W.E. Dietrich and M.J. Selby, 
%Modelling fluvial erosion on regional to continental scales, 
{\it J. Geophys. Res.},   
99(B7), 13,971-13,986, (1994).

\bibitem{R94}
R. Rigon, A. Rinaldo, I. Rodriguez-Iturbe, 
%On landscape self-organization,  
{\it J. Geophys. Res.}, {\bf 99}, 11971-11985, (1994).

\bibitem{B87}
P. Bak, C. Tang, and K. Wiesenfeld, 
%Self-organized criticality: an explanation of $1/f$ noise, 
{\it Phys. Rev. Lett.}, {\bf 59}, 381-384, (1987). 

\bibitem{AR95}
A. Rinaldo, W.E. Dietrich, G.K. Vogel, R. Rigon and I. Rodriguez-Iturbe, 
%Geomophological signatures of varying climate, 
{\it Nature}, 374, 632-636, (1995).

\bibitem{H57}
J.T. Hack, 
%Studies of longitudinal profiles in Virginia and Maryland,
{\it U.S. Geol. Surv. Prof. Paper  294-B},  1, (1957).

\bibitem{S96}
V. Sapozhnikov, and E. Foufoula-Georgiou, 
%Do current landscape evolution  models show self-organized criticality?, 
{\it Water Resour. Res.}, 32(4), 1109-1112, (1996).

\bibitem{R96}
R. Rigon, I. Rodriguez-Iturbe, A. Giacometti, A. Maritan, D. Tarboton,  
A. Rinaldo, 
%On Hack's law, 
{\it Water Resour. Res.}, {\bf32}, 3367-3374, (1996).  

\bibitem{RI92b}
I. Rodriguez-Iturbe, A. Rinaldo, R. Rigon, R. L. Bras,
A. Marani e E.J. Ijjasz-Vasquez, 
%Energy dissipation, runoff production, and the 3-dimensional structure of river basins,
{\it Water Resour. Res}, (28)4, 1095-1103, (1992).

\bibitem{RI92c}
I. Rodriguez-Iturbe, A. Rinaldo, 
R. Rigon, R. L. Bras, E. Ijjasz-Vasquez, 
%Fractal structures as least energy patterns: the case of river networks, 
{\it Geophys. Res. Lett.} {\bf 19}, 889-893, (1992).

\bibitem{AR92}
A. Rinaldo, I. Rodriguez-Iturbe, I., 
R. Rigon, R. L. Bras, E. Ijjasz-Vasquez, 
A. Marani, 
%Minimum energy and fractal structures of drainage networks, 
{\it Water Resour. Res.} {\bf 28}, 2183-2190, (1992).

\bibitem{SB96}
K. Sinclair, R.C. Ball
{\it Phys. Rev. Lett.} {\bf 76} 3360-3363, (1996).

\bibitem{data}
Some colour pictures relative at the evolution of models
A,B,C,D are available at
http://pil.phys.uniroma1.it/\~{ }gcalda/river.html

\end{thebibliography}
\end{document}